\documentclass[12pt]{article}
\usepackage{amsmath}
\usepackage{graphicx}
\usepackage{bm}
\usepackage{fancyhdr}
\usepackage{amssymb}
\usepackage{setspace}
\usepackage{epsfig}
\usepackage{overpic}
\usepackage{caption,subcaption}
\usepackage{epstopdf}
\begin{document}



\def\a{\alpha}
\def\b{\beta}
\def\d{\delta}
\def\e{\epsilon}
\def\g{\gamma}
\def\h{\mathfrak{h}}
\def\k{\kappa}
\def\l{\lambda}
\def\o{\omega}
\def\p{\wp}
\def\r{\rho}
\def\t{\tau}
\def\s{\sigma}
\def\z{\zeta}
\def\x{\xi}
\def\V={{{\bf\rm{V}}}}
 \def\A{{\cal{A}}}
 \def\B{{\cal{B}}}
 \def\C{{\cal{C}}}
 \def\D{{\cal{D}}}
\def\K{{\cal{K}}}
\def\O{\Omega}
\def\R{\bar{R}}
\def\T{{\cal{T}}}
\def\L{\Lambda}
\def\f{E_{\tau,\eta}(sl_2)}
\def\E{E_{\tau,\eta}(sl_n)}
\def\Zb{\mathbb{Z}}
\def\Cb{\mathbb{C}}

\def\R{\overline{R}}

\def\beq{\begin{equation}}
\def\eeq{\end{equation}}
\def\bea{\begin{eqnarray}}
\def\eea{\end{eqnarray}}
\def\ba{\begin{array}}
\def\ea{\end{array}}
\def\no{\nonumber}
\def\le{\langle}
\def\re{\rangle}
\def\lt{\left}
\def\rt{\right}

\newtheorem{Theorem}{Theorem}
\newtheorem{Definition}{Definition}
\newtheorem{Proposition}{Proposition}
\newtheorem{Lemma}{Lemma}
\newtheorem{Corollary}{Corollary}
\newcommand{\proof}[1]{{\bf Proof. }
        #1\begin{flushright}$\Box$\end{flushright}}

\baselineskip=20pt

\newfont{\elevenmib}{cmmib10 scaled\magstep1}
\newcommand{\preprint}{
   \begin{flushleft}
   \end{flushleft}\vspace{-1.3cm}
   \begin{flushright}\normalsize
   \end{flushright}}
\newcommand{\Title}[1]{{\baselineskip=26pt
   \begin{center} \Large \bf #1 \\ \ \\ \end{center}}}
\newcommand{\Author}{\begin{center}
   \large \bf
Yi Qiao${}^{a,b}$, Zhirong Xin${}^{a,b}$, Kun Hao${}^{a,b}$, Junpeng Cao${}^{c,d,e}$, Wen-Li Yang${}^{a,b,f}\footnote{Corresponding author:
wlyang@nwu.edu.cn}$, Kangjie Shi${}^{a,b}$ and ~Yupeng Wang${}^{c,d,e}\footnote{Corresponding author: yupeng@iphy.ac.cn}$

 \end{center}}
\newcommand{\Address}{\begin{center}

     ${}^a$Institute of Modern Physics, Northwest University,
     Xian 710127, China\\
     ${}^b$Shaanxi Key Laboratory for Theoretical Physics Frontiers,  Xian 710127, China\\
     ${}^c$Beijing National Laboratory for Condensed Matter
           Physics, Institute of Physics, Chinese Academy of Sciences, Beijing
           100190, China\\
     ${}^d$Collaborative Innovation Center of Quantum Matter, Beijing,
     China\\
     ${}^e$School of Physical Sciences, University of Chinese Academy of Sciences, Beijing, China\\
     ${}^f$School of Physics, Northwest University, Xi'an 710127, China

   \end{center}}

\preprint \thispagestyle{empty}
\bigskip\bigskip\bigskip

\Title{Twisted boundary energy and low energy excitation of the XXZ spin torus at the ferromagnetic region} \Author

\Address \vspace{0.1cm}

\vspace{1truecm}

\begin{abstract}

We investigate the thermodynamic limit of the one-dimensional ferromagnetic XXZ model with twisted (or antiperiodic ) boundary condition. It is shown that the distribution of the
Bethe roots of the inhomogeneous Bethe Ansatz equations (BAEs) for the ground state as well as for the low-lying excited states satisfy the string hypothesis,
although the inhomogeneous BAEs are not in the standard product form which has made the study of the corresponding thermodynamic limit nontrivial.
We also obtain the twisted boundary energy induced by the non-trivial
twisted boundary conditions in the thermodynamic limit.
\vspace{1truecm}


\end{abstract}

\newpage

\hbadness=10000

\tolerance=10000

\hfuzz=150pt

\vfuzz=150pt
\section{Introduction}
\setcounter{equation}{0}
The XXZ spin-$\frac{1}{2}$ torus model is the XXZ spin chain with  twisted (or antiperiodic) boundary condition \cite{Yun95,Bat95,Nie09,Nic13}, and it is tightly related
to the recent study on the boundary states of matter. In the works \cite{Yun95,Bat95}, Baxter's ``pair propagation through a vertex" was used and BA-like equations were derived and solved for the gapped anti-ferromagnetic regime. Based on the BA solutions, the interfacial tension are also obtained \cite{Bat95}.
In the critical regime, some very interesting results about the low-lying excitations are calculated and conformal field theory signatures are
derived by Niekamp, Wirth and Frahm \cite{Nie09}.
By means of the separation of variables, the BA-like equations and form factors are obtained \cite{Nic13}.
The integrability of this model is associated with  the six-vertex R-matrix \cite{Yan67,Yan68,Bax72,Bax82}. Due to the twisted boundary condition, the $U(1)$-symmetry is broken, making it much different from the periodic XXZ chain. For an example, the lack of an obvious reference state prevents us from applying the conventional Bethe Ansatz methods \cite{Yan661,Yan662,Yan663,Skl78,Fad81,Kor93,Tak99} to solve the model.

The off-diagonal Bethe Ansatz (ODBA) method is a newly developed analytic theory to compute exact solutions of quantum integrable models, especially for those with nontrivial integrable boundaries \cite{Wan13,Cao13,Cao14,Nep13}. Exact solution of such a system is characterized by an inhomogeneous $T-Q$ relations where the Bethe roots should satisfy the inhomogeneous Bethe Ansatz equations (BAEs).
Then, a natural question is what is the distribution of the Bethe roots of the inhomogeneous BAEs in the complex plane. It is very important because it is the start point to study the thermodynamic properties of the system.
However, due to the existence of the inhomogeneous term, it is hard to use the usual thermodynamic Bethe Ansatz method \cite{Tak99}. Some interesting approaches are proposed \cite{Li142,Wen17}. For examples, because the inhomogeneous BAEs can degenerate into the conventional ones when the model parameters taking some special values, then one can use the results at the degenerate points to get the actual values \cite{Li142}. Another method is that one can study the contribution of the inhomogeneous term, then the thermodynamic limit can be obtained by using the finite-size scaling behavior \cite{Wen17, Wen17-1}. In this paper, we directly investigate the Bethe roots from the inhomogeneous BAEs without any approximation. We use the XXZ spin-$\frac{1}{2}$ torus as the example. From the numerical analysis, we obtain the structure of the Bethe roots for the ground state as well as for the low-lying excited states in the thermodynamic limit. Based on them, the physical quantities such as the twisted boundary energy and the gap are studied.

The paper is organized as follows. Section 2 serves as an introduction of the model and its exact solution. In section 3, we discuss the distribution of Bethe roots of the ground state. Base on them, we calculate the twisted boundary energy. In section 4, the nearly degenerate states are studied. In section 5, we discuss the elementary excitation and the energy gap. In section 6, the limiting behavior is considered, which is used to check our results.
Concluding remarks are given in section 7.

\section{Spin-$\frac{1}{2}$ XXZ torus}
\setcounter{equation}{0}
The spin-$\frac{1}{2}$ XXZ torus is characterized by the Hamiltonian
\bea\label{Ham}
H=-\sum_{j=1}^N\big[ \sigma_j^x \sigma_{j+1}^x + \sigma_j^y \sigma_{j+1}^y +\cosh\eta \sigma_j^z \sigma_{j+1}^z \big],
\eea
where $N$ is the number of sites, $\eta$ is the crossing parameter (or anisotropic parameter)  and the boundary condition is the twisted one, namely,
\bea
\sigma_{N+1}^\alpha=\sigma_1^x\sigma_1^\alpha\sigma_1^x,\,\,{\rm for}\,\, \alpha=x, y, z.\label{Anti-periodic}
\eea
It is remarked that the twisted boundary condition (\ref{Anti-periodic}) breaks the bulk $U(1)$-symmetry (c.f. the spin-$\frac{1}{2}$ XXZ chain with the periodic
boundary condition: $\sigma_{N+1}^\alpha=\sigma_1^\alpha$).

The integrability of the model is associated with the well-known six-vertex $R$-matrix
\begin{eqnarray}\label{R-matrix}
  R_{0,j}(u)= \frac{1}{2} \left[ \frac{\sinh(u+\eta)}{\sinh \eta} (1+\sigma^z_j \sigma^z_0) +\frac{\sinh u}{\sinh \eta} (1- \sigma^z_j \sigma^z_0)   \right]+ \frac{1}{2} (\sigma^x_j \sigma^x_0 +\sigma^y_j \sigma^y_0) ,
\end{eqnarray}
where $u$ is the spectral parameter. From the $R$-matrix, we can define the monodromy matrix as
\begin{equation}\label{monodromy-matrix}
  T_0(u)=\sigma_0^x R_{0,N}(u-\theta_N) \cdots R_{0,1}(u-\theta_1)=\left(
                                                                     \begin{array}{cc}
                                                                       C(u) & D(u)\\
                                                                        A(u)& B(u)\\
                                                                     \end{array}
                                                                   \right).
\end{equation}
The $R$-matrix and the monodromy matrix satisfy the RTT relation
\begin{equation}
  R_{0,\bar{0} }(u-v) T_0(u)T_{\bar{0}}(v)=T_{\bar{0}}(v) T_0(u) R_{0,\bar{0} }(u-v).
\end{equation}
The transfer matrix of the system is defined as
\begin{equation}
t(u)=tr_0 T_0(u)=B(u)+C(u).
\end{equation}
From the RTT relation, one can prove that
\begin{equation}
  [t(u),t(v)]=0,
\end{equation}
thus the system is integrable. The transfer matrix can generate all the conserved quantities and the Hamiltonian (\ref{Ham}) is chosen as the first order derivative of the logarithm of the transfer matrix
\begin{equation}
  H=-2 \sinh \eta  \frac{\partial \ln t(u)}{\partial u}\big|_{u=0, \{ \theta_j=0 \}} + N \cosh \eta.
\end{equation}

The Hamiltonian (\ref{Ham}) can be exactly solved by using the ODBA method \cite{Wan13, Cao13}. The eigen-energy is then expressed in terms of the Bethe roots
\bea\label{energy}
E=2\,i\sinh\eta\sum_{j=1}^N \big[ \cot\big(u_j+\frac{i\eta}{2}\big)-\cot\big(u_j-\frac{i\eta}{2}\big) \big]
-N\cosh\eta-2\sinh\eta,
\eea
where the Bethe roots $\{u_j\}$ should satisfy the inhomogeneous BAEs
\bea\label{BAEs2}
&&e^{iu_j}\prod_{l=1}^N\frac{\sin(u_j-u_l+i\eta)}{\sin(u_j+\frac{1}{2}i\eta)}=
e^{-iu_j}\prod_{l=1}^N\frac{\sin(u_j-u_l-i\eta)}{\sin(u_j-\frac{1}{2}i\eta)}+
2i\,e^{-\frac{1}{2}N\eta}\sin\big(u_j-\sum_{l=1}^N{u_l}\big), \no
\\[5pt]
&&\quad \quad j=1,\cdots,N. \quad \quad \quad
\eea
We note that the period of Bethe roots is $\pi$, thus we fix the real part of Bethe roots in the interval $[-\frac{\pi}{2}, \frac{\pi}{2})$.

\section{Bethe roots for the ground state}
\setcounter{equation}{0}

\begin{figure}[htbp]
\centering
\includegraphics[height=6cm,width=8cm]{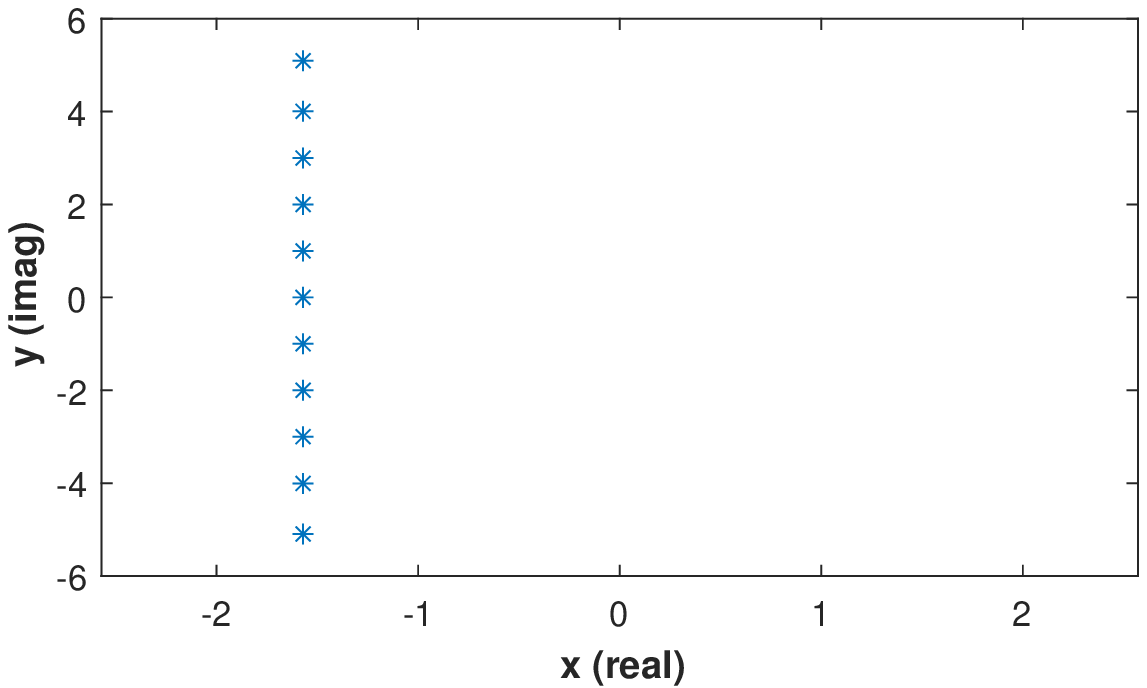}
\caption{The distribution of Bethe roots of the inhomogeneous BAEs for the ground state, where $\eta=1$ and $N=11$. The string is located at the boundary of the period ($-\frac{\pi}{2}$).
The difference among the imaginary part of Bethe roots is equal to $\eta i$.}\label{GS_Betheroots}
\end{figure}

Here, we consider the case that $\eta$ is real and solve the inhomogeneous BAEs (\ref{BAEs2}) numerically. The values of Bethe roots for the ground states for finite system-size are listed in Table \ref{gsbr-eta0.5-c1} and \ref{gsbr-eta1-c1}. From the data, we find that the real part of Bethe roots is nearly $- \pi/2$, and the difference between the imaginary part of two Bethe roots is nearly constant and the value is $i\eta$.
Thus the Bethe roots form a single string \cite{Tak99}. In order to see this point clearly, we draw them in Figure \ref{GS_Betheroots}. From it we see that the string is located at the boundary of the period ($-\frac{\pi}{2}$).

We should mention that for the odd $N$ case, there exists another $N$-string which locates at the imaginary axis. These two kinds of strings give the same ground state energy thus the corresponding Bethe states are degenerate.
\begin{table}[!ht]
\caption{ Bethe roots of the inhomogeneous BAEs for the ground state, where $\eta=0.5$. }\label{gsbr-eta0.5-c1}
\scriptsize
\centering
\begin{tabular}{|c|c|c|c|c|c|}
\hline {$\{u_i\}$$\setminus$N}& $ 7 $  & $ 8 $ & $ 9 $ & $ 10 $ & $ 11 $ \\ \hline
 $u_1$ &  $-1.5708+1.4373i$ & $-1.5708+1.6820i$ & $-1.5708+1.9671i$ & $-1.5708+2.2153i$ &  $-1.5708+2.4755i$ \\
 $u_2$ &  $-1.5708+1.2052i$& $-1.5708+1.4370i$ & $-1.5708+1.6317i$ & $-1.5708+1.8656i$ &  $-1.5708+2.0966i$ \\
 $u_3$ &  $-1.5708+0.5208i$& $-1.5708+0.7628i$ & $-1.5708+1.0133i$ & $-1.5708+1.2572i$ &  $-1.5708+1.5054i$ \\
 $u_4$ &  $-1.5708-0.0000i$& $-1.5708+0.2624i$ & $-1.5708+0.5059i$ & $-1.5708+0.7533i$ &  $-1.5708+1.0013i$ \\
 $u_5$ &  $-1.5708-0.5208i$& $-1.5708-0.2624i$ & $-1.5708-0.0000i$ & $-1.5708+0.2503i$ &  $-1.5708+0.5002i$ \\
 $u_6$ &  $-1.5708-1.2052i$& $-1.5708-0.7628i$ & $-1.5708-0.5059i$ & $-1.5708-0.2503i$ &  $-1.5708+0.0000i$ \\
 $u_7$ &  $-1.5708-1.4373i$& $-1.5708-1.4370i$ & $-1.5708-1.0133i$ & $-1.5708-0.7533i$ &  $-1.5708-0.5002i$ \\
 $u_8$ & $ $ & $-1.5708-1.6820i$ & $-1.5708-1.6317i$ & $-1.5708-1.2572i$ &  $-1.5708-1.0013i$ \\
 $u_9$ & $ $ & $ $ & $-1.5708-1.9671i$ & $-1.5708-1.8656i$ &  $-1.5708-1.5054i$ \\
 $u_{10}$ & $ $ & $ $ & $ $ & $-1.5708-2.2153i$ &  $-1.5708-2.0966i$ \\
 $u_{11}$ & $ $ & $ $ & $ $ & $ $ &  $-1.5708-2.4755i$ \\
\hline\end{tabular}
\end{table}
\begin{table}[!ht]
\caption{Bethe roots of the inhomogeneous BAEs for the ground state, where $\eta=1$. }\label{gsbr-eta1-c1}
\scriptsize
\centering
\begin{tabular}{|c|c|c|c|c|c|}
\hline {$\{u_i\}$$\setminus$N} & $ 7 $& $ 8 $ & $ 9 $ & $ 10 $ & $ 11 $ \\ \hline
 $u_1$ &  $-1.5708+3.0989i$& $-1.5708+3.5966i$ & $-1.5708+4.0958i$ & $-1.5708+4.5954i$ &  $-1.5708+5.0953i$ \\
 $u_2$ &  $-1.5708+2.0087i$& $-1.5708+2.5072i$ & $-1.5708+3.0066i$ & $-1.5708+3.5064i$ &  $-1.5708+4.0063i$ \\
 $u_3$ &  $-1.5708+1.0003i$ & $-1.5708+1.5001i$ & $-1.5708+2.0001i$ & $-1.5708+2.5000i$ &  $-1.5708+3.0000i$ \\
 $u_4$ &  $-1.5708-0.0000i$ & $-1.5708+0.5000i$ & $-1.5708+1.0000i$ & $-1.5708+1.5000i$ &  $-1.5708+2.0000i$ \\
 $u_5$ &  $-1.5708-1.0003i$& $-1.5708-0.5000i$ & $-1.5708-0.0000i$ & $-1.5708+0.5000i$ &  $-1.5708+1.0000i$ \\
 $u_6$ &  $-1.5708-2.0087i$& $-1.5708-1.5001i$ & $-1.5708-1.0000i$ & $-1.5708-0.5000i$ &  $-1.5708-0.0000i$ \\
 $u_7$ &  $-1.5708-3.0989i$& $-1.5708-2.5072i$ & $-1.5708-2.0001i$ & $-1.5708-1.5000i$ &  $-1.5708-1.0000i$ \\
 $u_8$ &  $ $& $-1.5708-3.5966i$ & $-1.5708-3.0066i$ & $-1.5708-2.5000i$ &  $-1.5708-2.0000i$ \\
 $u_9$ &  $ $& $ $ & $-1.5708-4.0958i$ & $-1.5708-3.5064i$ &  $-1.5708-3.0000i$ \\
 $u_{10}$ &  $ $& $ $ & $ $ & $-1.5708-4.5954i$ &  $-1.5708-4.0063i$ \\
 $u_{11}$ &  $ $& $ $ & $ $ & $ $ &  $-1.5708-5.0953i$ \\
\hline\end{tabular}
\end{table}

Based on above facts, we conclude that all the Bethe roots may form the string solution for the ground state
\bea\label{string-GS}
u_j=-\frac{\pi}{2}+\big(\frac{N+1}{2}-j\big)\eta i +o(N), \quad j=1,\cdots,N,
\eea
where $o(N)$ stands for a small correction which is related with $N$, and $i$ is the imaginary unit.
Substituting the string hypothesis (\ref{string-GS}) into the energy expression (\ref{energy}) and neglecting the small correction, we obtain the ground state energy
\bea\label{E_stringGE}
E_{0}=-N\cosh\eta-2\sinh\eta+4\sinh\eta \tanh\big(\frac{N\eta}{2}\big).
\eea

\begin{figure}[!htbp]
\centering
\includegraphics[height=6cm,width=12cm]{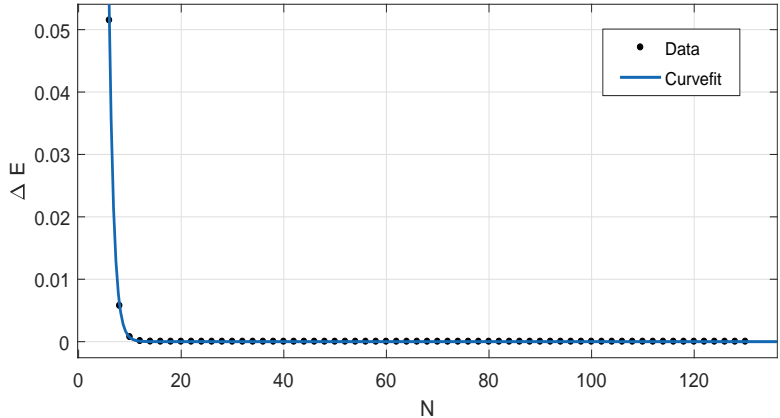}
\caption{The relative error to the ground state energy at $\eta=1$. Here $\Delta E= E_{0}- E_{0}^{DMRG}$ and $E_{0}^{DMRG}$ is the ground state energy calculated from DMRG method. The data can be fitted as $\Delta E=35.51e^{-1.089N} $. We can easily find the relative error tends to zero rapidly when $N$ tends to infinity. }\label{DMRG_8_2_130}
\end{figure}

In order to check the validity of string hypothesis (\ref{string-GS}), we calculate the ground state energy of the system by exactly diagonalizing the Hamiltonian (\ref{Ham}) up to $N=19$ and compare the results with those obtained by the equation (\ref{E_stringGE}). The data is listed in Table \ref{gsbr-etqqqqqa1-c1}. We see that the analytical and numerical results agree with each other very well.
\begin{table}[!ht]
\centering
\caption{The comparison of ground state energies. Here $\Delta E=E_{0}-E_{0}^{ED}$, $E_{0}$ is analytic results and $E_{0}^{ED}$ is the one calculated by the exact diagonalization. Despite some data with small $N$, the data can be fitted as $\Delta E \propto e^{-\alpha N}, \alpha \approx \eta$. }\label{gsbr-etqqqqqa1-c1}
\footnotesize
\begin{tabular}{|c|c|c|c|c|c|}
\hline {N$\setminus$$\Delta E$$\setminus$$\eta$} & $0.5 $ & $1 $ & $1.5 $ & $2 $ \\ \hline
 $2$ & $-0.3342142165$ & $0.1435417877$ & $0.7458761805$ &  $1.2074616511$ \\
 $3$ & $-0.0952609137$ & $0.3716389985$ & $0.5043105875$ &  $0.4167008182$ \\
 $4$ & $0.0509806795$ & $0.2848861255$ & $0.1707344774$ &  $0.0704866355$ \\
 $5$ & $0.1197282891$ & $0.1407699247$ & $0.0381958000$ &  $0.0089281224$ \\
 $6$ & $0.1334305287$ & $0.0514910722$ & $0.0077697666$ &  $0.0011529804$ \\
 $7$ & $0.1141588186$ & $0.0169264962$ & $0.0016597727$ &  $0.0001544483$ \\
 $8$ & $0.0810290178$ & $0.0057499892$ & $0.0003658926$ &  $0.0000208680$ \\
 $9$ & $0.0491663635$ & $0.0020416802$ & $0.0000813935$ &  $0.0000028234$ \\
 $10$ & $0.0269446294$ & $0.0007405552$ & $0.0000181472$ &  $0.0000003821$ \\
 $11$ & $0.0144539470$ & $0.0002708820$ & $0.0000040484$ &  $0.0000000517$ \\
 $12$ & $0.0079734142$ & $0.0000994183$ & $0.0000009033$ &  $0.0000000070$ \\
 $13$ & $0.0045457565$ & $0.0000365387$ & $0.0000002015$ &  $0.0000000009$ \\
 $14$ & $0.0026510859$ & $0.0000134366$ & $0.0000000450$ &  $0.0000000001$ \\
 $15$ & $0.0015679356$ & $0.0000049423$ & $0.0000000100$ &  $0.0000000000$ \\
 $16$ & $0.0009355173$ & $0.0000018180$ & $0.0000000022$ &  $0.0000000000$ \\
 $17$ & $0.0005613593$ & $0.0000006688$ & $0.0000000005$ &  $0.0000000000$ \\
 $18$ & $0.0003380990$ & $0.0000002460$ & $0.0000000001$ &  $0.0000000000$ \\
 $19$ & $0.0002041302$ & $0.0000000905$ & $0.0000000000$ &  $0.0000000000$ \\
\hline\end{tabular}
\end{table}
For the larger system-size, we also check the validity of (\ref{string-GS}) and (\ref{E_stringGE}) by the density matrix renormalization group (DMRG) method. The result is shown in Figure \ref{DMRG_8_2_130}.
Then, we can conclude that the Bethe roots for the ground state form the string solution in the thermodynamic limit and Eq. (\ref{E_stringGE}) gives the energy of the system.

Now, we calculate the twisted boundary energy. It is well-known that the ground state energy of the XXZ spin chain with periodic boundary condition is \cite{Tak99}
\bea\label{E_periodGE}
E_{0}^p=-N\cosh\eta.
\eea
We define the twisted boundary energy as
\bea \label{boundary1}
E_{t}&=&E_{0}-E_{0}^p.
\eea
Substituting Eqs.(\ref{E_stringGE}) and (\ref{E_periodGE}) into (\ref{boundary1}), we obtain
\bea \label{boundary111}
E_{t}=-2\sinh\eta+4\sinh\eta \tanh\big(\frac{N\eta}{2}\big).
\eea
In the thermodynamic limit, the twisted boundary energy arrives at
\bea
E_{t}=2\sinh\eta. \label{boundary2}
\eea

\section{Nearly degenerate states}
\setcounter{equation}{0}

From the numerical calculation, we also find that the XXZ spin torus model has some nearly degenerate states with the ground state, which have following properties:

1. They are almost degenerate states with exponentially small gaps for the finite system-size. As shown in Figure 3, the energy difference between the ground state and nearly degenerate states satisfies the law $\Delta E \propto e^{ -\beta N }, \beta \approx \eta$, which means that the difference will exponentially tends to zero with the increasing $N$. Thus in the thermodynamic limit, the nearly degenerate states are degenerate to the ground state.
\begin{figure}[!htbp]
\begin{minipage}{0.30\linewidth}
  \leftline{\includegraphics[width=8.5cm]{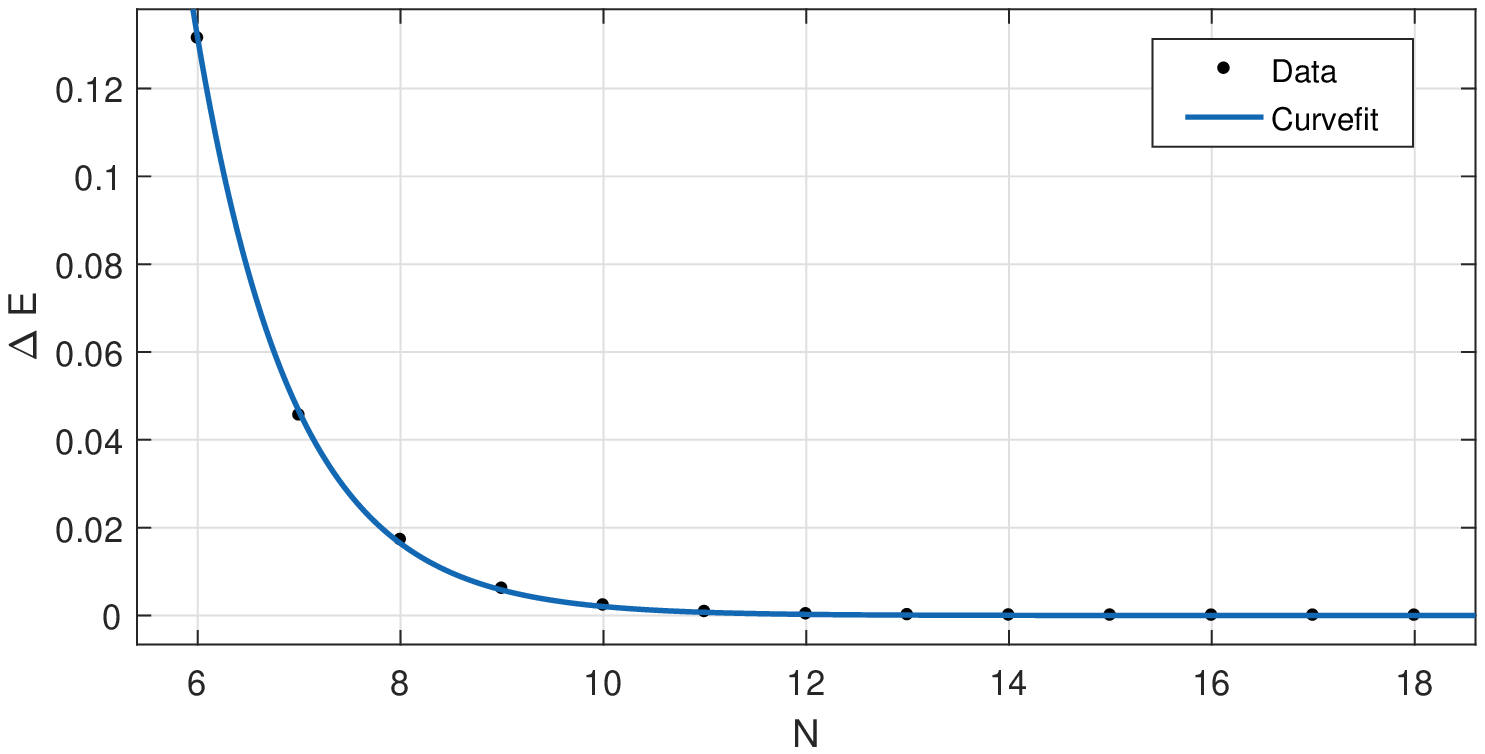}}
  \rightline{(a) $\eta=1$\quad}
\end{minipage}
\hfill
\begin{minipage}{0.30\linewidth}
  \rightline{\includegraphics[width=8.5cm]{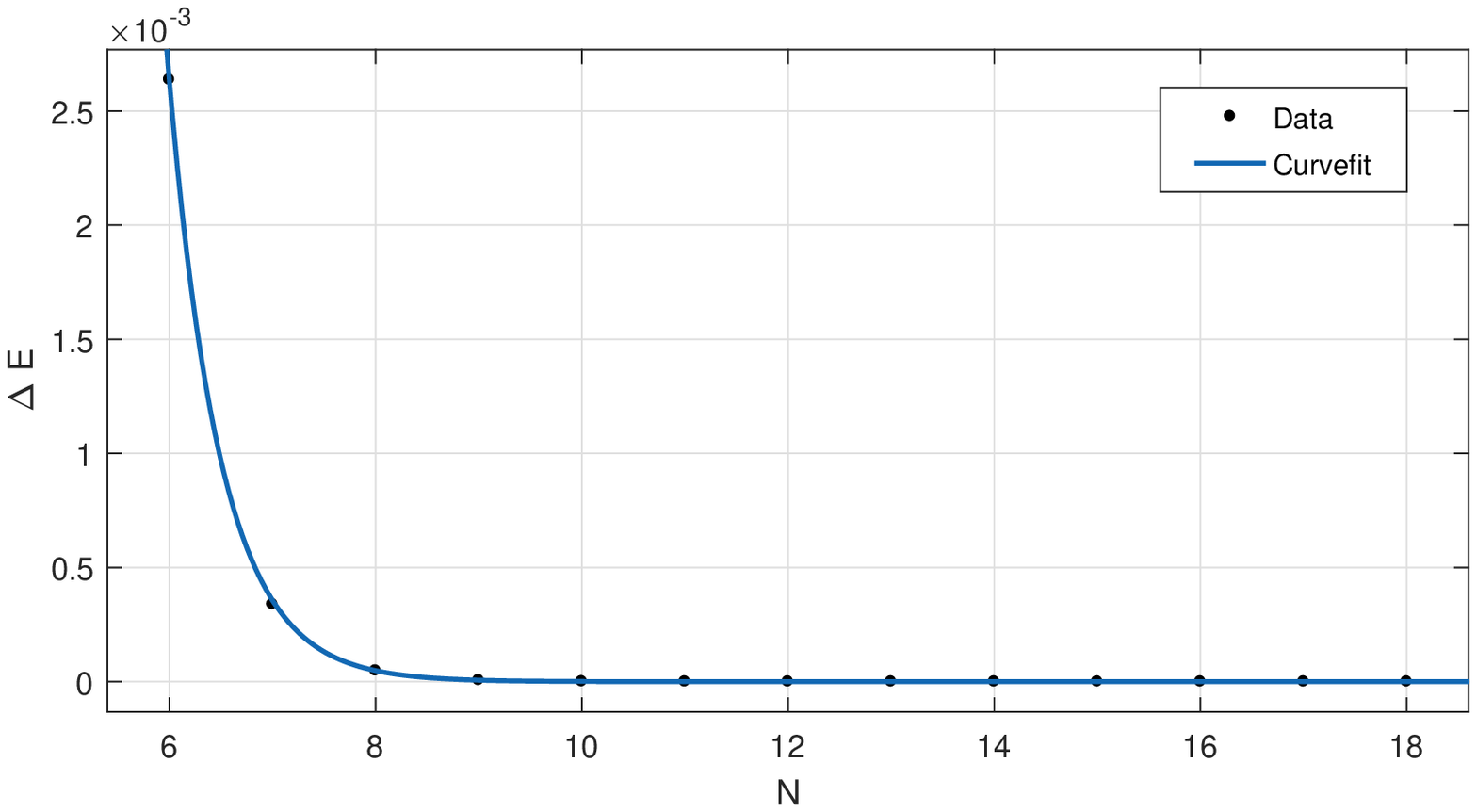}}
  \leftline{\qquad (b) $\eta=2$}
\end{minipage}
\caption{The energy difference between the ground state and nearly degenerate states. Here $\Delta E = E_{nmax}^{ED}-E_{0}^{ED}$, and and $E_{nmax}^{ED}$ is the highest energy in the nearly degenerate states calculated by the exact diagonalization, $E_{0}^{ED}$ is the ground state energy calculated by the exact diagonalization. The data can be fitted as (a) $\Delta E = 67.56e^{-1.041N}$ (b) $\Delta E = 427.4e^{-1.999N}$. From the fitting, we see that the energy difference tends to zero rapidly with the increasing system-size. }\label{NDE_diff}
\end{figure}

2. The number of the nearly degenerate states is $2N-2$. In the case of $\eta \rightarrow +\infty$, the model degenerates to a Ising-like model. The system possesses $Z_2$ symmetry and the boundary conditions at different site number are equivalent. It is easy to find that the degeneracy of the ground state is $2N$, which is self-consistent with the numerical results. Based on this fact, we can obtain the following physical picture. When the crossing parameter $\eta$ tends to infinity, the main contribution in Hamiltonian (\ref{Ham}) to the energy comes from the $\sigma_j^z \sigma_{j+1}^z$ terms, and the degeneracy of the ground state is $2N$. When the $\eta$ is smaller, the rest parts of Hamiltonian make contributions and the ground state splits into several separate energies, corresponding to the ground state and the nearly degenerate states here. Because of the $Z_2$ symmetry, the degeneracy of the ground state is 2, so we have $2N-2$ nearly degenerate states.

\begin{figure}[!htbp]
\begin{minipage}{0.30\linewidth}
  \rightline{\includegraphics[width=3.5cm]{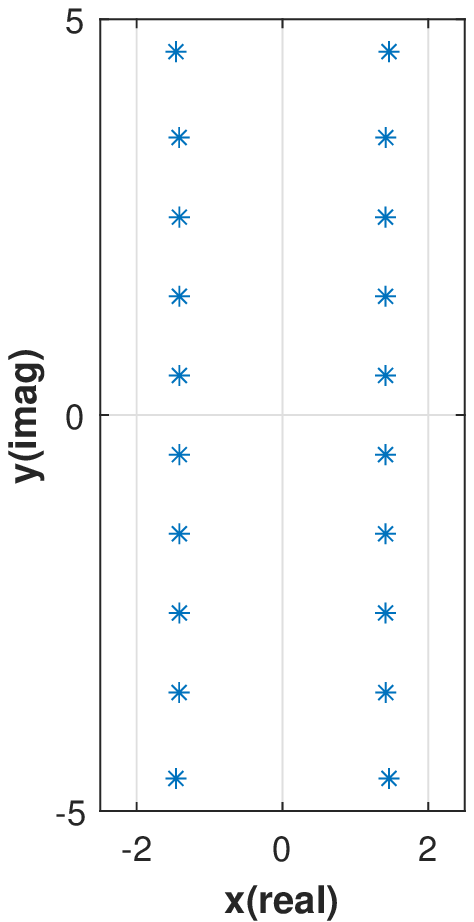}}
  \rightline{(a)$E_{n1}$\qquad \ \ }
\end{minipage}
\hfill
\begin{minipage}{0.30\linewidth}
  \centerline{\includegraphics[width=3.5cm]{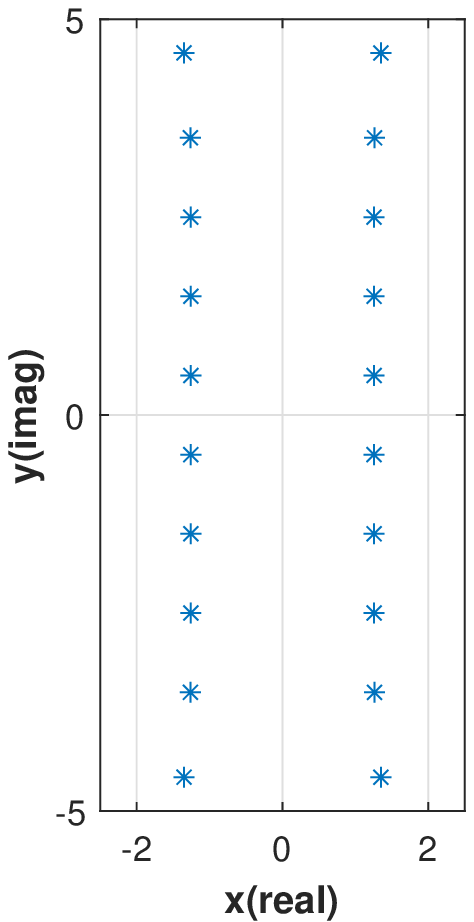}}
  \centerline{\quad  (b)$E_{n2}$}
\end{minipage}
\hfill
\begin{minipage}{0.30\linewidth}
  \leftline{\includegraphics[width=3.5cm]{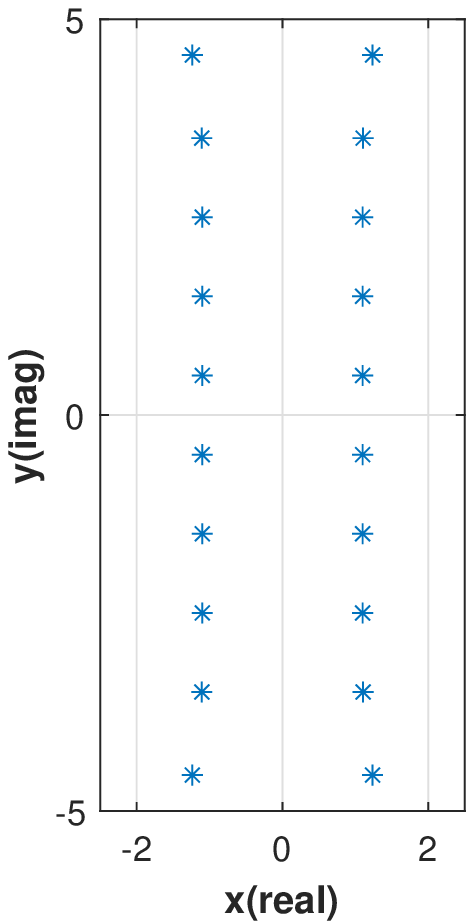}}
  \leftline{ \qquad \quad (c)$E_{n3}$}
\end{minipage}
\caption{The distributions of Bethe roots for the nearly degenerate states, where $\eta=1$ and $N=10$. (a), (b) and (c) stand for 3 different nearly degenerate energies $E_{n1}$, $E_{n2}$ and $E_{n3}$, respectively. The string structure can be seen very clearly.}\label{NDS_Betheroots}
\end{figure}

3. As shown in Figure \ref{NDS_Betheroots}, the Bethe roots for the nearly degenerate states also form a $N$-string structure, but the position is moved and the deviation $o$ is larger than that for the ground state.


\section{Elementary excitation}
\setcounter{equation}{0}

\begin{figure}[!htbp]
\centering
\includegraphics[height=6cm,width=9cm]{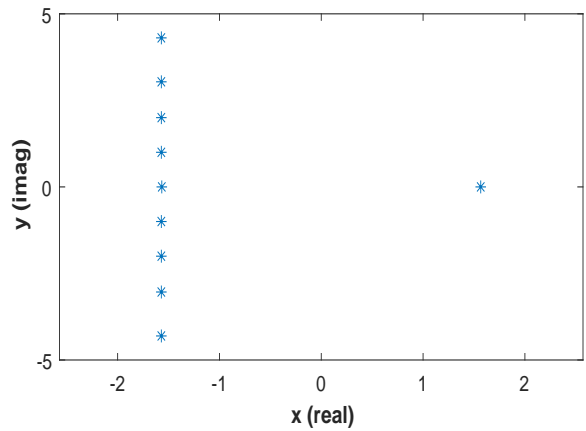}
\caption{Elementary excitation state Bethe roots distribution of $\eta=1, N=10$.}\label{EE_Betheroots}
\end{figure}

\begin{figure}[!ht]
\centering
\includegraphics[height=6cm,width=12cm]{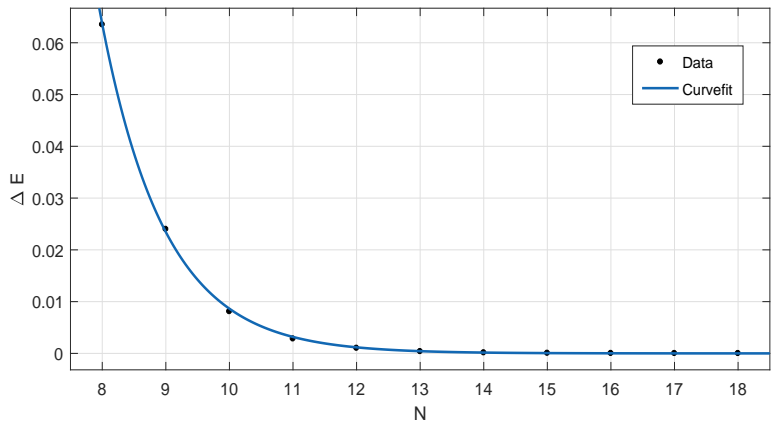}
\caption{The energy difference between the theoretical value and the actual one. Here $\eta=1$, $\Delta E = E_{1}-E_{1}^{ED}$, and $E_{1}^{ED}$ is the lowest excitation energy calculated by the exact diagonalization. The data can be fitted as $\Delta E = 187.3e^{-0.9985N}$. From the fitting, we see that the energy difference tends to zero rapidly with the increasing system-size. }\label{EE_diff_eta1}
\end{figure}

Now, we consider the elementary excitation of the XXZ spin torus model. As shown in Figure \ref{EE_Betheroots}, we find that the distribution of Bethe roots for the lowest excited states can be described by a $(N-1)$-string plus an additional real root. Meanwhile, the string and the real root are nearly located at the interval boundary $-\frac{\pi}{2}$ \footnote{we regard $\frac{\pi}{2}$ as the same point with $-\frac{\pi}{2}$ due to the periodicity.}.
After some more precision calculation, we conclude that the Bethe roots for the lowest excited state take the form of
\bea
u_1=-\frac{\pi}{2}+o_1(N);\quad u_j=-\frac{\pi}{2}+\big(\frac{N}{2}-j+1\big)\eta i +o_2(N);\quad j=2,\cdots,N, \label{EE_Bethero11111111ots}
\eea
where $o_1(N)$ and $o_2(N)$ stand for the small deviations and $i$ is the imaginary unit. The energy corresponding to this kind of excitation is
\bea\label{E_stringEE}
E_{1}=-N\cosh\eta-2\sinh\eta+4\sinh\eta \tanh\left[\frac{(N-1)\eta}{2}\right]+4\sinh\eta \tanh\frac{\eta}{2}.
\eea
We also check the validity of Eq. (\ref{E_stringEE}) by the exact diagonalization and the result is shown in Figure \ref{EE_diff_eta1}.
Again, we see that with the increasing $N$, the energy difference between the analytic result (\ref{E_stringEE}) and the actual values tends to zero rapidly.
Thus the string solution (\ref{EE_Bethero11111111ots}) is correct in the thermodynamic limit.

The energy gap of the XXZ spin torus is defined as
\bea\label{anti-gap1}
E_{gap}&=&E_{1}-E_{0} \no \\
&=&4\sinh\eta \tanh\frac{\eta}{2} + 4\sinh\eta \tanh\left[\frac{(N-1)\eta}{2}\right]-4\sinh\eta \tanh\big(\frac{N\eta}{2}\big).
\eea
In the thermodynamic limit, the energy gap reads
\bea\label{anti-gap2}
E_{gap}=4\sinh\eta \tanh\frac{\eta}{2},
\eea
which is the same as that of the XXZ spin chain with periodic boundary condition.

\section{Limiting behavior}
\setcounter{equation}{0}

In order to check above results again, now, we consider some limit case of $\eta \rightarrow +\infty$.
In this case, the Ising-like spin coupling $\sigma_j^z \sigma_{j+1}^z$ in Hamiltonian (\ref{Ham}) is dominated and the model degenerates into the one-dimensional Ising model with a twisted boundary condition. Evidently, the ferromagnetic state is the ground state and the corresponding ground state energy is known as
\bea
\frac{E_{0}^{Ising}}{\cosh\eta}=-N+2.
\eea
Taking the $\eta \rightarrow +\infty$ limit of Eq. (\ref{E_stringGE}), we have
\bea
\frac{E_{0}\left. \right|_{\eta=+\infty}}{\cosh\eta}=-N+2,
\eea
which is consistent with the result of Ising model.
The lowest excited energy of Ising model with a twisted boundary condition is
\bea
\frac{E_{1}^{Ising}}{\cosh\eta}=-N+6.
\eea
Taking the $\eta \rightarrow +\infty$ limit of Eq. (\ref{E_stringEE}), we have
\bea
\frac{E_{1}\left. \right|_{\eta=+\infty}}{\cosh\eta}=-N+6.
\eea
Therefore, the results derived from the string hypothesis (\ref{string-GS}) and (\ref{EE_Bethero11111111ots}) are consistent with the already known results.

\section{Conclusions}
In this paper, we have studied the thermodynamic limit of
the one-dimensional ferromagnetic XXZ model with the twisted (or antiperiodic) boundary condition which is described by the Hamiltonian (\ref{Ham}) and (\ref{Anti-periodic}). It is shown that even for the inhomogeneous BAEs (\ref{BAEs2}), the  corresponding roots for the ground state appear to form a string (\ref{string-GS}). This fact enables us to calculate the twisted boundary energy of the model given by (\ref{boundary2}). By using the similar method, we further investigate the elementary excitation and obtain the energy gap of the model which is the same as that of the periodic boundary condition case.

\section*{Acknowledgments}

The financial supports from the National Program
for Basic Research of MOST (Grant Nos. 2016YFA0300600 and
2016YFA0302104), the National Natural Science Foundation of China
(Grant Nos. 11434013, 11425522, 11547045, 11774397, 11775178 and 11775177), the Major Basic Research Program of Natural Science of Shaanxi Province
(Grant Nos. 2017KCT-12, 2017ZDJC-32) and the Strategic Priority Research Program of the Chinese
Academy of Sciences, and the Double First-Class University Construction Project of Northwest University are gratefully acknowledged.
Y. Qiao and Z. Xin are also partially supported by the NWU graduate student innovation funds (Grant Nos. YZZ15088, YYB17003).

\end{document}